\documentclass[journal,twocolumn]{IEEEtran}
\usepackage{amsmath}

\newcommand{\figref}[1]{{Fig.}~\ref{#1}}
\newcommand{\tabref}[1]{{Table}~\ref{#1}}

\newcommand{\sref}[1]{{Section}~\ref{#1}}

\def\bb0{{\mathbb{0}}}

\def\bb{{\mathbf{b}}}

\def\bff{{\mathbf{f}}}

\def\bh{{\mathbf{h}}}

\def\b0{{\mathbf{0}}}

\def\bbE{{\mathbb{E}}}

\def\sf0{{\mathsf{0}}}

\usepackage[table]{xcolor}
\usepackage{epstopdf}
\usepackage{makecell}
\usepackage{array,colortbl,multirow,multicol,booktabs,ctable}
\usepackage{enumerate}
\usepackage{algorithmicx}
\usepackage{algorithm}
\usepackage{amsmath}

\usepackage{amsfonts}
\usepackage[noend]{algpseudocode}
\usepackage{float}
\usepackage{soul}
\usepackage{hyperref}
\usepackage{makeidx}
\usepackage{bbm}
\usepackage{graphicx}
\usepackage[font=small]{caption}
\usepackage{array}
\usepackage{epsfig}
\usepackage{comment}
\usepackage{textcomp}
\usepackage{subcaption}
\usepackage{nopageno}
\usepackage{cite}
\usepackage{amssymb}

\begin{document}
	\bstctlcite{IEEEexample:BSTcontrol} % Forces the use of et al in bibliography
	\title{LiDAR-Aided Mobile Blockage Prediction in Real-World Millimeter Wave Systems}
	
	\author{\IEEEauthorblockN{Shunyao Wu, Chaitali Chakrabarti, and Ahmed Alkhateeb}\\
		\vspace{5pt}
		\IEEEauthorblockA{\textit{School of Electrical, Computer and Energy Engineering}\\
			\textit{Arizona State University - Email: \{vincentw, chaitali, alkhateeb\}@asu.edu} \thanks{This work is supported in part by the National Science
					Foundation under Grant No. 2048021.}}\\
				}
	
	\maketitle
	
	\begin{abstract}
		Line-of-sight link blockages represent a key challenge for the reliability and latency of millimeter wave (mmWave) and terahertz (THz) communication networks. This paper proposes to leverage LiDAR sensory data to provide awareness about the communication environment and proactively predict dynamic link blockages before they happen. This allows the network to make proactive decisions for hand-off/beam switching which enhances its reliability and latency. We formulate the LiDAR-aided blockage prediction problem and present the first real-world demonstration for LiDAR-aided blockage prediction in mmWave systems. In particular, we construct a large-scale real-world dataset, based on the DeepSense 6G structure, that comprises co-existing LiDAR and mmWave communication measurements in  outdoor vehicular scenarios. Then, we develop an efficient LiDAR data denoising (static cluster removal) algorithm and a machine learning model that  proactively predicts dynamic link blockages. Based on the real-world dataset, our LiDAR-aided approach is shown to achieve 95\% accuracy in predicting blockages happening within 100ms and more than 80\% prediction accuracy for blockages happening within one second. If used for proactive hand-off, the proposed solutions can potentially provide an order of magnitude saving in the network latency, which highlights a promising direction for addressing the blockage challenges in mmWave/sub-THz networks.  
		
	\end{abstract}
	
	\begin{IEEEkeywords}
		Millimeter wave, LiDAR, blockage prediction.
	\end{IEEEkeywords}
	\thispagestyle{empty}
	
	% \newpage

	\section{Introduction} 
	\label{sec:Intro}

	Millimeter wave (mmWave) and sub-terahertz (sub-THz) communication systems have the potential of providing high data rates thanks to the large bandwidth available at these frequency bands \cite{rappaport2019wireless,wang2018millimeter}.  Realizing these gains in mobile networks, however, requires overcoming the key challenge represented by the sensitivity of the high-frequency signals to blockages \cite{Andrews2017}. These blockages may abruptly disconnect the line-of-sight (LOS) links, which critically affects the reliability and latency of the mobile networks. Current solutions for the mmWave/THz blockage and reliability problems are mainly focused on multiple connectivity where the  user simultaneously keeps multiple links to one or more basestations \cite{polese2017improved,petrov2017dynamic}. This, however, inefficiently consumes the network resources and does not completely solve the reliability/latency challenges as the switching between links is still performed in a reactive way (after a link failure is observed). 
	
	% Prior work on blockage prediction 
	A promising approach for overcoming the link blockage problems in mmWave/THz networks is to leverage machine/deep learning solutions to proactively predict the blockages before they happen \cite{alkhateeb2018machine}. This allows the network to act proactively in switching the user to a different link/basestation. In \cite{alkhateeb2018machine}, recurrent neural networks were leveraged to predict future blockages based on sequences of previously used beams. Further, \cite{alrabeiah2020deep} leveraged sub-6GHz channels to identify/predict close-to-happen mmWave blockages. The approaches in \cite{alkhateeb2018machine,alrabeiah2020deep}, are mainly suitable for predicting stationary blockages or blockages that are a few milliseconds ahead. To predict future blockages hundreds of milliseconds before they happen, leveraging other sensory data such as cameras \cite{charan2021vision, Alrabeiah_Cameras, Alrabeiah_URLL, Alrabeiah_ViWi} was proposed. Deploying cameras, however, may not always be possible for privacy/regulatory reasons. This motivates the research for using other data modalities for future blockage prediction.

	In this paper, we propose to leverage LiDAR sensory data to provide information about the communication environment and help proactively predict LOS mmWave link blockages potentially seconds before they happen. The main contributions of the paper can be summarized as follows:
	\begin{itemize}
		\item We formulate the LiDAR-aided blockage prediction problem and develop an efficient convolutional neural network (CNN)-based model to proactively predict moving link blockages using LiDAR data.
		
		\item We construct a framework for collecting co-existing mmWave and LiDAR data, and use it to build a large-scale real-world outdoor dataset ($\sim$ 125 thousand data points). The dataset consists of mmWave beam training data,  LiDAR sensory data,  and the corresponding images, and can be used to study multiple problems including the LiDAR-aided blockage prediction task. 
		
		\item We develop a static cluster removal (pre-processing) algorithm for the data generated by low-cost LiDAR sensors. The approach denoises this data and accurately extracts the traces of the moving objects, thus improving the performance of our machine learning models.
		
	\end{itemize}
	
	Using the collected real-world dataset, our approach achieves 95\% accuracy for predicting blockages happening within a future prediction interval of 0.1s and 80\% for a prediction interval of 1s. Thanks to proactively predicting blockages, our solution can reduce the hand-off latency from 222.8ms (with reactive hand-off) to 35.43ms, highlighting a promising solution for  mmWave/sub-THz networks.

		\begin{figure*}[t]
		\centering
		\includegraphics[width=1\linewidth]{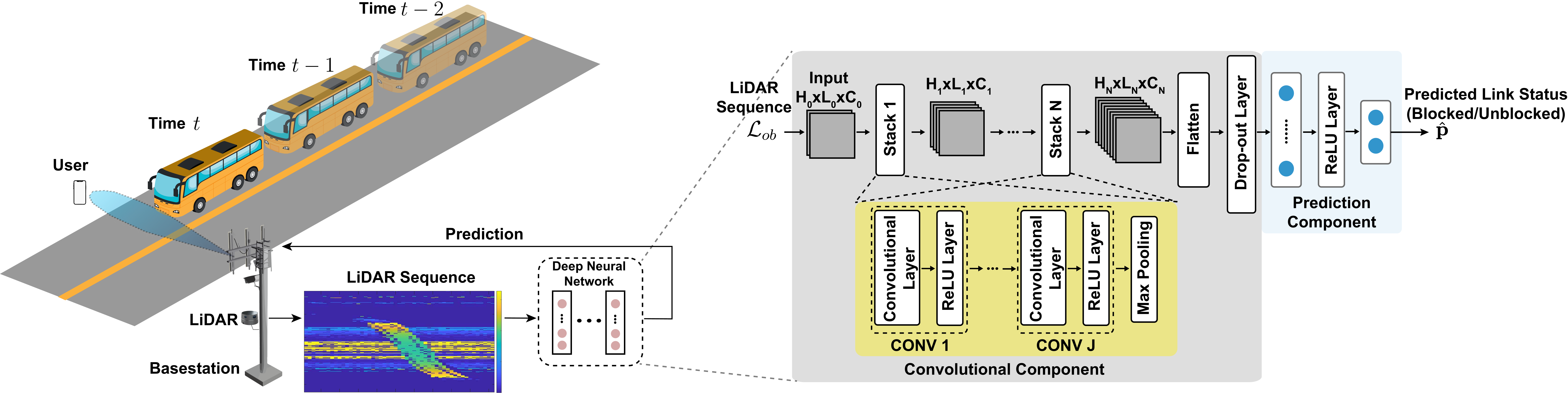}
		\caption{Illustration of the overall system model where a mmWave/sub-THz basestation leverages a LiDAR sensor to provide environment awareness and enable the proposed proactive link blockage prediction approach. The figure also depicts the adopted neural network model which  consists  of convolutional and prediction components. }
		\label{fig:system diagram}
	\end{figure*}

	\section{System and Channel Model} \label{sec:sys_model}

	We adopt a mmWave communication system where a basestation with $M_\mathrm{A}$-element antenna array is used to serve a stationary user. The basestation is further equipped with a LiDAR sensor to provide awareness about the surrounding environment and moving scatterers/blockages, as shown in \figref{fig:system diagram}. The basestation employs a pre-defined beamsteering codebook of $M$ beams, $\boldsymbol{\mathcal F} = \{\mathbf f_m\}_{m=1}^{M}$, where $\mathbf f_m$ is a beamsteering vector that directs the signal towards direction $\theta_m= \theta_\text{offset}+ \text{FoV}/M$, with FoV denoting the field of view of the wireless beamforming system \cite{Zhang_RL}. In our testbed, described in \sref{sec:testbed}, we consider a phased array with $M_\mathrm{A}$=16 elements and a  codebook of $M=64$ beamforming vectors, with steering angles uniformly quantize the range $[-\pi/4, \pi/4]$. It is worth mentioning here, though, that the proposed blockage prediction approaches in this paper can be applied to more general array architectures.
	
	To account for the variations of the channel over time, we adopt a block fading channel model where the channel is assumed to be constant over a time duration of $\tau_B$. Further, we adopt an OFDM signal transmission model of $K$ subcarriers. We define $\bh_k[t] \in \mathbb C^{M\times 1}$ as the downlink channel from the base station to the user at the $k$-th subcarrier for discrete time instance $t$, where $t\in \mathbb Z$.  At time $t$, if the beamsteering vector $\bff_m$ is adopted by the basestation for the downlink transmission, then the received signal at subcarrier $k$ is 
	\begin{equation}\label{sig_model}
	r_{k,m}[t] = \bh_k[t]^T \bff_m s_k[t] + n_k[t]
	\end{equation}
	where $s_k[t]$ is the transmitted symbol at the $k$-th subcarrier and $t$-th time instance, $\bbE{\left|s_k[t]\right|^2}=1$, and $n_k[t] \sim \mathcal {CN}(0,\sigma_n^2)$ is a noise sample.
	
	\begin{figure}[tb]
		\centering
		\includegraphics[width=1\columnwidth]{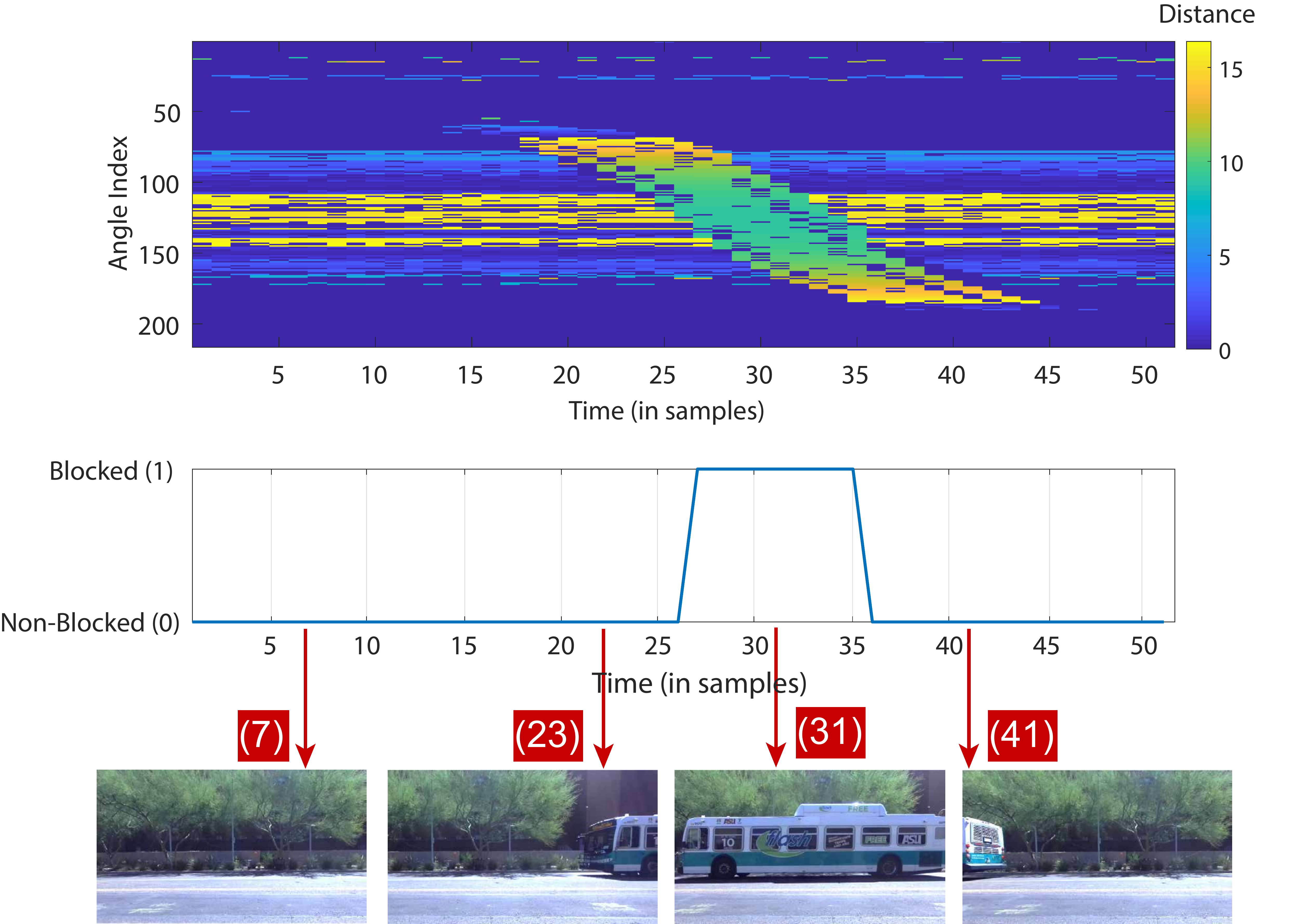}
		\caption{An example of the LiDAR pre-blockage signature: The upper plot shows the distance heatmap as a function of the angle and time instance. The middle plot depicts the corresponding link status and the bottom plot shows the image captured by the camera.}
		\label{fig:lidar_exp}
	\end{figure}

	\section{Problem Definition} \label{sec:prob_form}

	We investigate the potential of LiDAR sensors to proactively predict if the link between the mmWave base station and the user is going to be blocked in the near future. 
	
	Let $x[t]\in{\{0, 1\}}$ be the link status (blocked or unblocked) at time instance $t$. Further, we assume that at $t$, the LiDAR sensor provides sensory data $\mathbf L[t] \in \mathbb{R}^{P \times 2}$,  where $P$ represents the number of points collected by the  LiDAR sensor at time $t$ for a 360 degree scan; data at each point consists of an angle and a distance value representing the measured depth at this angle. We define $\mathcal{L}_{ob} $ as the sequence of LiDAR samples at the $T_{ob}$ previous time instances (observation window),  $\mathcal{L}_{ob} = \{ \mathbf L[t+n] \}_{n=-T_{ob}+1}^{0}$.  
	
	Now, given the observed LiDAR sequence $\mathcal L_{ob}$, our task is to predict whether or not a link blockage will occur within a future time interval of $T_P$ instances. We use $b_{T_P}$ to indicate whether there is a blockage occurrence within that future interval or not. More formally, $b_{T_p}$ can be defined as follows:
	\vspace*{-0.05in}
	\begin{equation} \label{equ:p1_label}
	b_{T_P} = 
	\begin{cases}
	0, &  x[t+n_{p}] = 0 \quad \forall n_{p} \in \{ 1,\dots,T_P \} \\
	1, & \text{otherwise}
	\end{cases}       
	\end{equation}
	
	\vspace*{-0.05in}
	\noindent where $0$ indicates the absence of blockage and $1$ is the occurrence of a blockage. Thus, given the sequence $\mathcal{L}_{ob}$, our machine learning task is to predict the future blockage status $b_{T_p}$ with the highest accuracy. If $\hat {b}_{T_p}$ denotes the predicted link status, \textbf{our goal is to maximize the successful blockage prediction probability} $\mathbb P_1( \hat {b}_{T_p}= b_{T_p} | \mathcal L_{ob})$. 
	
	%\vspace*{-0.05in}
	
	\section{Proactive Blockage Prediction using LiDAR Pre-Blockage Signature} \label{sec:lidar_sign}
	A typical LiDAR sensor sends pulsed light waves into the surrounding environment. These pulses are reflected by the objects and returned to the sensor; and the sensor uses the round-trip time to calculate the distance it traveled. By sending and receiving laser beams, the LiDAR sensor collects a 2-D point cloud map of the surrounding environment. We propose to leverage these LiDAR sensory data to detect if an object is going to block the mmWave communication  link between the basestation and the mobile user. \figref{fig:lidar_exp} shows an example when a communication link is getting blocked by a moving object. The heatmap on the top represents the sensed distance for every quantized angle (direction) level as a function of time. The horizontal lines represent static objects (since their distances from the LiDAR device do not change over time). The middle plot shows the link status and the bottom photos show the corresponding scenario.  Note that from time instance 13 to 26, the color of the pattern changes from yellow to green, indicating that the distance between the moving object (bus) and the LiDAR sensor has become shorter. This matches the scenario as the bus was approaching. The color becomes yellow again after time instance 35 implying that the object is moving away from the LiDAR sensor. \textbf{Thus, as the blockage approaches the link, we can see a clear pre-blockage pattern in the LiDAR heatmap that can potentially be leveraged for proactive blockage prediction.}
	
	%\section{Machine Learning Model} \label{ML_model}
	To learn this pre-blockage signature at the basestation, we design the CNN model depicted in \figref{fig:system diagram}. In this figure, $H$, $L$ and $C$ denote the number of rows, columns and channels of the input data. The CNN architecture consists of a convolutional component followed by a prediction component. In the convolutional component, the first stack (Stack 1) takes the input collected LiDAR data (i.e., $\mathcal S_{ob}$) whose dimension is $H_0 \times L_0 \times C_0$ and passes the output with the dimension $H_1 \times L_1 \times C_1$ to the next stack. We design $N$ similar stacks and the last stack (Stack $N$) is followed by a flatten layer (a flatten layer converts the 2-D matrix to 1-D vector). Each stack contains $J$ convolutional blocks, each of which consists of a convolutional layer and a ReLU layer, and a max-pooling layer that occurs at the end of the stack. The output of the convolutional component is fed to a fully connected (FC) layer followed by a classifier. The classifier outputs a probability vector ($\hat{\mathbf p}$) of whether the link status is blocked or not in $T_P$ future time instances.
	%which is illustrated in neural network model in \figref{fig:system diagram}.
	
	\section{Testbed and Scenarios} \label{sec:testbed}

	\begin{figure*}[tb]
		\centering
		\includegraphics[width=1\linewidth]{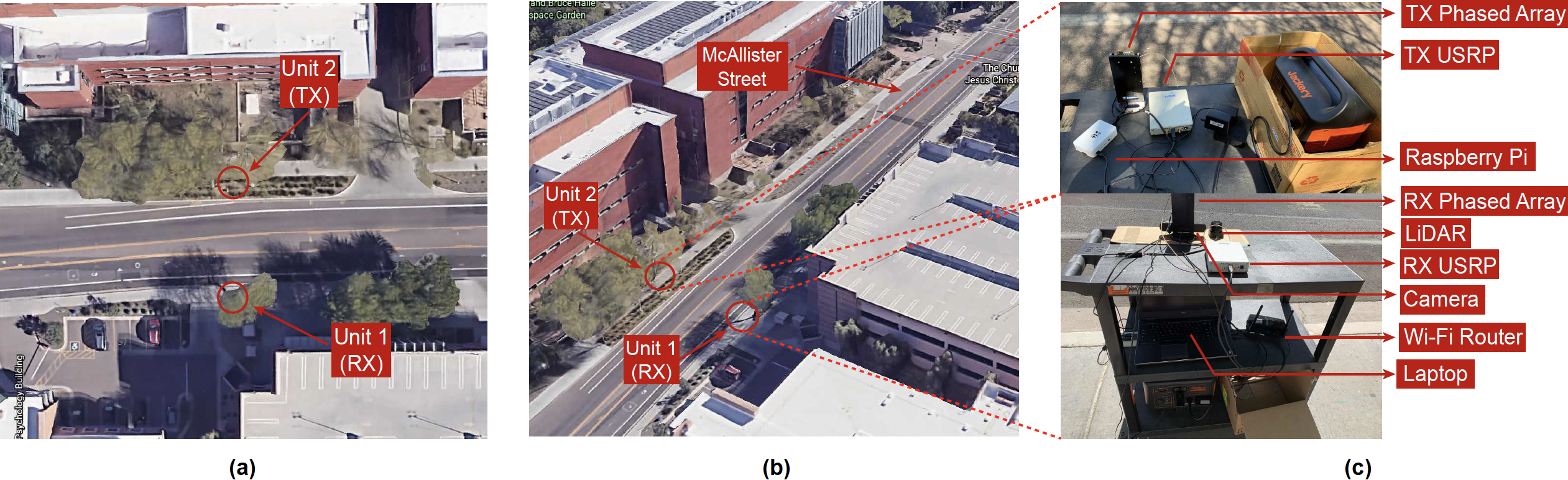}
		\caption{Illustration of the data collection scenario/location, where Unit 1 (TX) and Unit 2 (RX) were deployed at the two sides of the street. The right subfigure shows the equipment used in our dataset collection.}
		\label{fig:exp_setup}
	\end{figure*}

	To evaluate the performance of the proposed approach in real-world environments, we generate measurement-based datasets, following the footsteps of the DeepSense 6G dataset \cite{DeepSense}. We deploy a testbed in an outdoor wireless environment to collect real-world multimodal measurements and construct, what will be henceforth called, \textit{seed datasets}. 
	
	\subsection{Testbed Description} \label{subsec:testbed}
	The DeepSense 6G dataset framework \cite{DeepSense} defines a generic structure for sensing/communication datasets where a number of units, each equipped with a set of sensors, collect co-existing sensory/communication data. We adopt the \textbf{DeepSense Testbed 3} \cite{mmwave_journal} and add a synchronized LiDAR sensor. \figref{fig:exp_setup} shows the \textbf{DeepSense Testbed 3} which consists of two stationary units, namely Unit 1 and Unit 2. Unit 1 collects mmWave beam training measurements, visual data, and LiDAR data while Unit 2 is equipped with a mmWave transmitter. At each capture, the system collects a number of measurements including a LiDAR sample and an RGB image. The scanning range of the LiDAR is 16 meters and the motor spin frequency is 10Hz.

	\subsection{DeepSense Scenarios 24-27} \label{subsec:scenario}
	
	We collect data in an outdoor wireless environment representing a two-way city street, as shown in \figref{fig:exp_setup}. The two units of \textbf{DeepSense Testbed 3} are placed on the two sides of the street. The LiDAR at Unit 1 continuously scans the environment. The testbed collects data samples at a rate of 10 samples/s. Each data sample has multiple modalities including an RGB image and a LiDAR 360-degree point cloud, both collected by Unit 1. The important aspects of these DeepSense scenarios are summarized in \tabref{tbl:scen_20}.

	\begin{table}[t]
		\caption{Scenarios 24-27 (Multimodal Blockage I-IV)}
		\label{table}
		\centering
		\setlength{\tabcolsep}{5pt}
		\renewcommand{\arraystretch}{1.4}
		\begin{tabular}{|c|c|}
			\hline\hline
			\textbf{Testbed}             & 3               \\ \hline
			\textbf{Number of Instances} & \thead{Scenario 24: 11952 - Scenario 25: 35424 \\ Scenario 26: 36648 -Scenario 27: 39672\\ Combined:  123696 (from 3436 trajectories)   }                    \\ \hline
			\textbf{Number of Units}     & 2 \\ \hline
			\textbf{Data Modalities}     & \thead{RGB images, LiDAR point cloud, positions, \\ mmWave beam training measurements} \\ \hline \hline
			\multicolumn{2}{|c|}{\textbf{Unit 1: Stationary}} \\ \hline
			% 		\textbf{Type} & Stationary \\ \hline
			\textbf{Hardware elements} & \thead{RGB Camera, \\  mmWave receiver with 16-element \\ phased array, LiDAR} \\ \hline
			\textbf{Data Modalities} & \thead{RGB images, \\ LiDAR point cloud, GPS position, \\ mmWave beam training measurements}  \\
			\hline \hline
			\multicolumn{2}{|c|}{\textbf{Unit 2: Stationary}} \\ \hline
			% 		\textbf{Type} & Stationary \\ \hline
			\textbf{Hardware elements} &  \thead{mmWave transmitter with an omni-antenna }  \\ \hline
			%	\textbf{Container Number} & 2 \\ \hline
			\textbf{Data Modalities} & GPS position \\
			\hline\hline
		\end{tabular}
		\label{tbl:scen_20}
		%\vspace{-4mm}
	\end{table}

	\section{LiDAR Data Processing with \\ Static Cluster Removal (SCR)} \label{sec:SCR}
	In order to build the development dataset used in the LiDAR-aided blockage prediction ML task, the raw LiDAR data described in \sref{subsec:scenario} needs to be first pre-processed to (i) remove the noise created by static clusters and (ii) remove the sensory data collected from outside the communication field of view. To clarify the first point, we plot \figref{fig:SCR_object} and \figref{fig:SCR_ori} which show the moving object captured by the RGB camera and the corresponding point clouds of the raw LiDAR data. The points in the orange rectangular represent the trace of moving objects, the ones in the red circles represent the static clusters or objects in the scenario, while the points in the blue circles represent the noisy points that are not present in the real scenario, which are called distracting path reflection points (also static noise). We are interested in the trace of moving objects and so the static objects and the path reflection noise should be removed.

	The aim of pre-processing is mainly to eliminate the cluster static noise points. We first use a field of view based filtering to erase the LiDAR sensory data collected from directions outside the field of view of interest (\sref{subsec:field_filter}). Next, we use a dictionary based cluster removal method to remove the unnecessary clusters in \sref{subsec:scr}.

	\subsection{Field of View Based Filtering} \label{subsec:field_filter}
	Since the objects of interest are between the transmitter and receiver, any LiDAR-detected object on the other side of this communication link need to be filtered out so that it does not distract the blockage prediction model. Assuming that the LiDAR device collects $P$ samples at every time instance, and each LiDAR sample has 2 values, angle $\phi$ in radians and distance $d$ in meters, the set of samples is $\mathcal L^{(t)} = \{(\phi,d)_p\}_{p=1}^P$, $\mathcal L$ denotes the raw LiDAR dataset, $t$ is the index of time instance. We use the field of view filter to clean the points outside the range ${\Phi}_1$ to ${\Phi}_2$. In this paper, we define ${\Phi}_1 = -\pi/6$, ${\Phi}_2 = \pi$ based on real measurements. We choose $P = 460$ at each time instance, since our LiDAR sensor collects 460 samples for a 360 degree point cloud.

	\subsection{Dictionary Based Cluster Removal} \label{subsec:scr}
	After the field of view based filtering, some of the static clusters in the LiDAR point cloud are eliminated. We now develop a static cluster removal (SCR) method to remove the rest. The SCR method is implemented in five steps: i) sorting the LiDAR data by their angle; ii) quantizing these angles; iii) quantizing the distance of the LiDAR data; iv) generating a static cluster dictionary from samples that contain no moving objects, and v) eliminating the static clusters according to the constructed dictionary. The sorting and quantization are needed to establish the mapping between points at different time instances.

	\textbf{Step 1 Sorting:} Although the number of collected LiDAR samples at each time instance is the same, they are not ordered by either their angles or distances. So the first step is to sort the LiDAR samples by their angles. In the sorting process, we append the zero-distance points at the end of the sorted important points, since these zero-distance points cannot provide effective information.
	
	\textbf{Step 2 Angle Quantization:} After sorting, the angle is in ascending order. We uniformly quantize the angle space to ensure that the points in one time instance can be mapped to those in other time instances within the same quantization step. This establishes a relation between the points at different time instances. We define $Q$ as the total number of angle quantization levels and $q$ as the index of the angle quantization level. The quantization angle is from ${\Phi}_1$ to ${\Phi}_2$ and the step size is denoted as $\Delta \Phi$. If there are multiple points at the same quantization level, we choose the median index of the points whose angle lay in the same quantization level and discard others. In this paper, we choose $Q = 216$.

	\begin{figure*}[tb]
		\centering
		\begin{subfigure}[t]{0.25\textwidth}
			\centering
			\includegraphics[width=\linewidth]{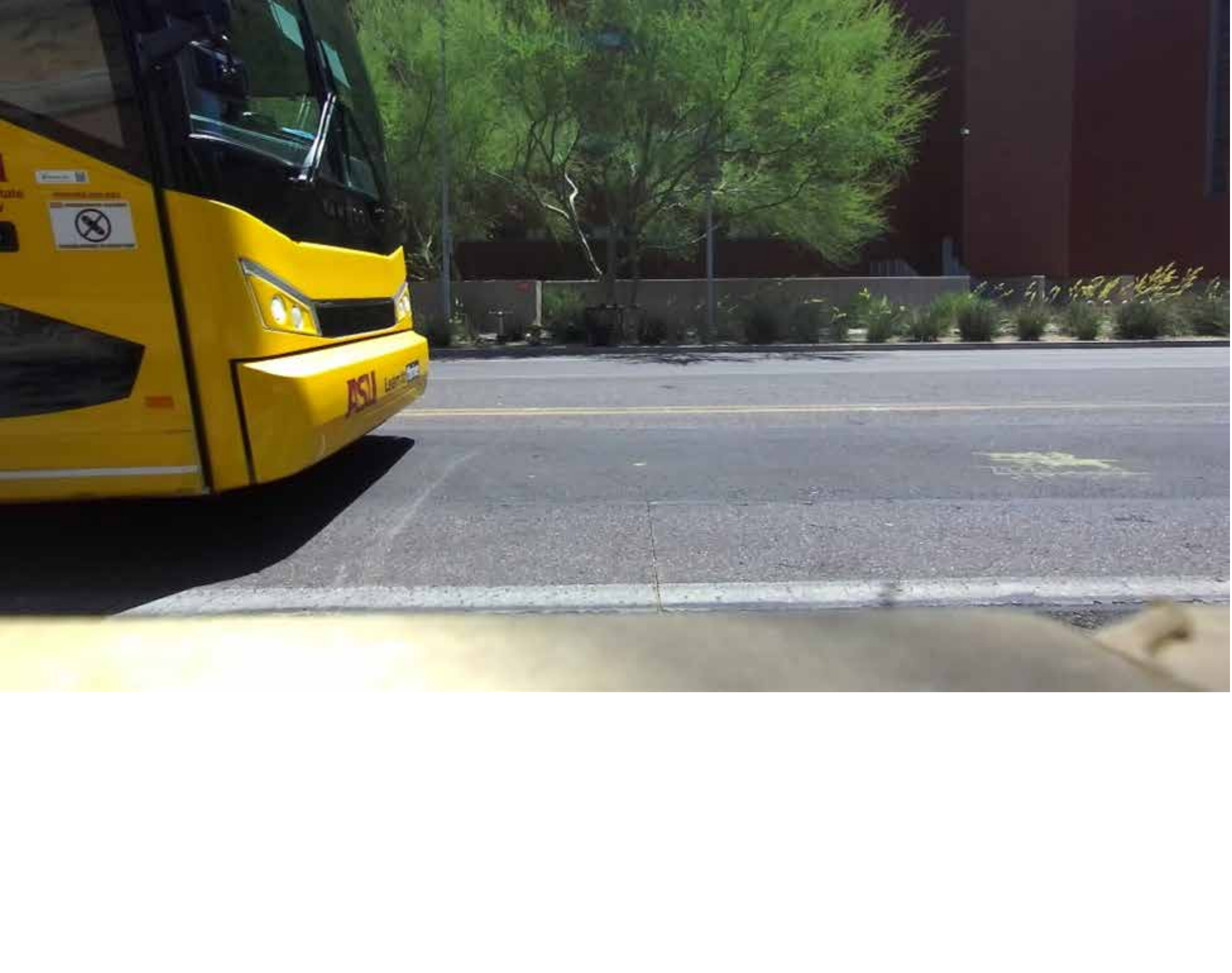}
			\caption{}
			\label{fig:SCR_object}
			% 	\vfill
		\end{subfigure}
		\begin{subfigure}[t]{0.35\textwidth}
			\centering
			\includegraphics[width=\linewidth]{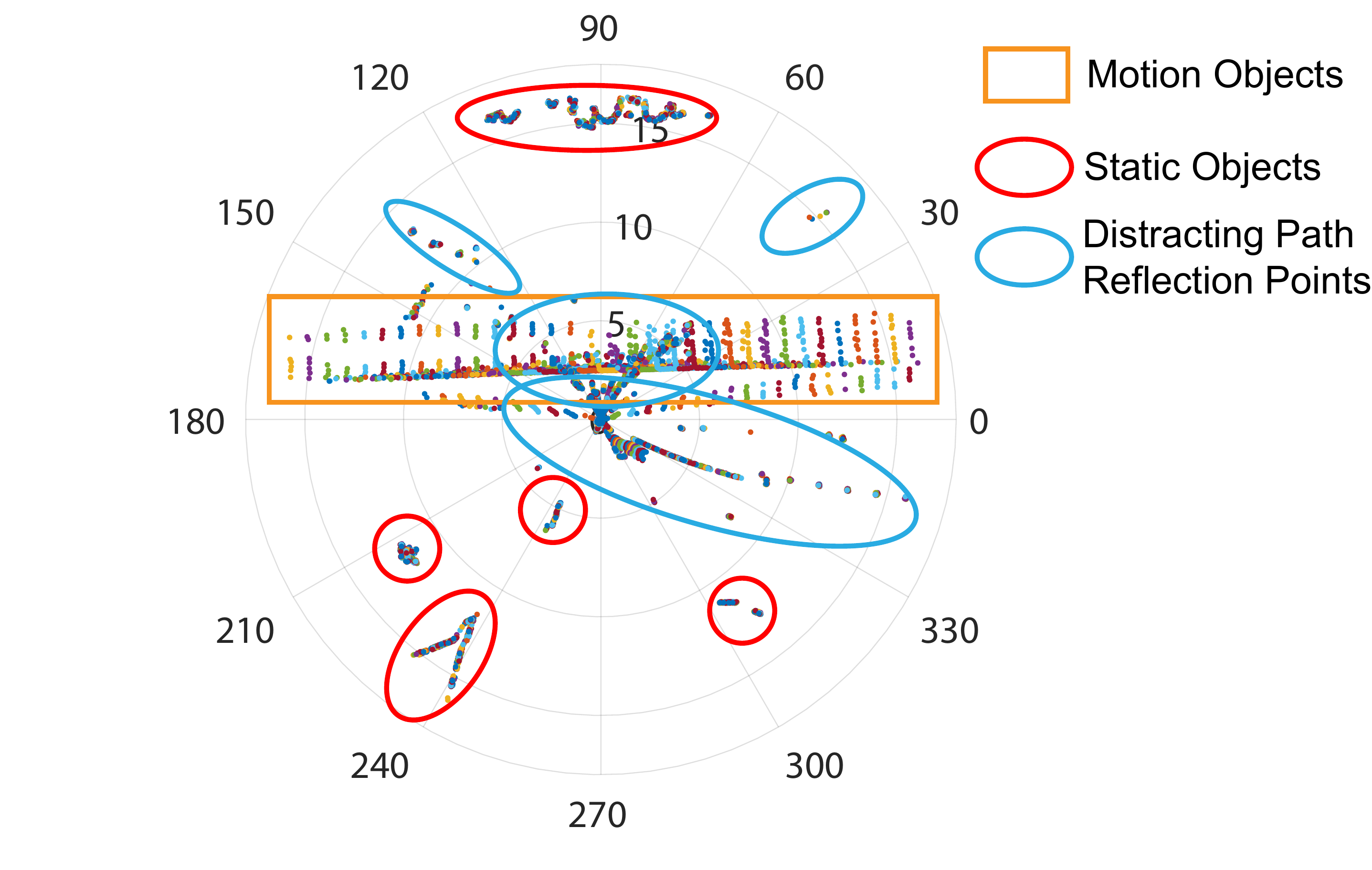}
			\caption{}
			\label{fig:SCR_ori}
		\end{subfigure}
		\begin{subfigure}[t]{0.35\textwidth}
			\centering
			\includegraphics[width=\linewidth]{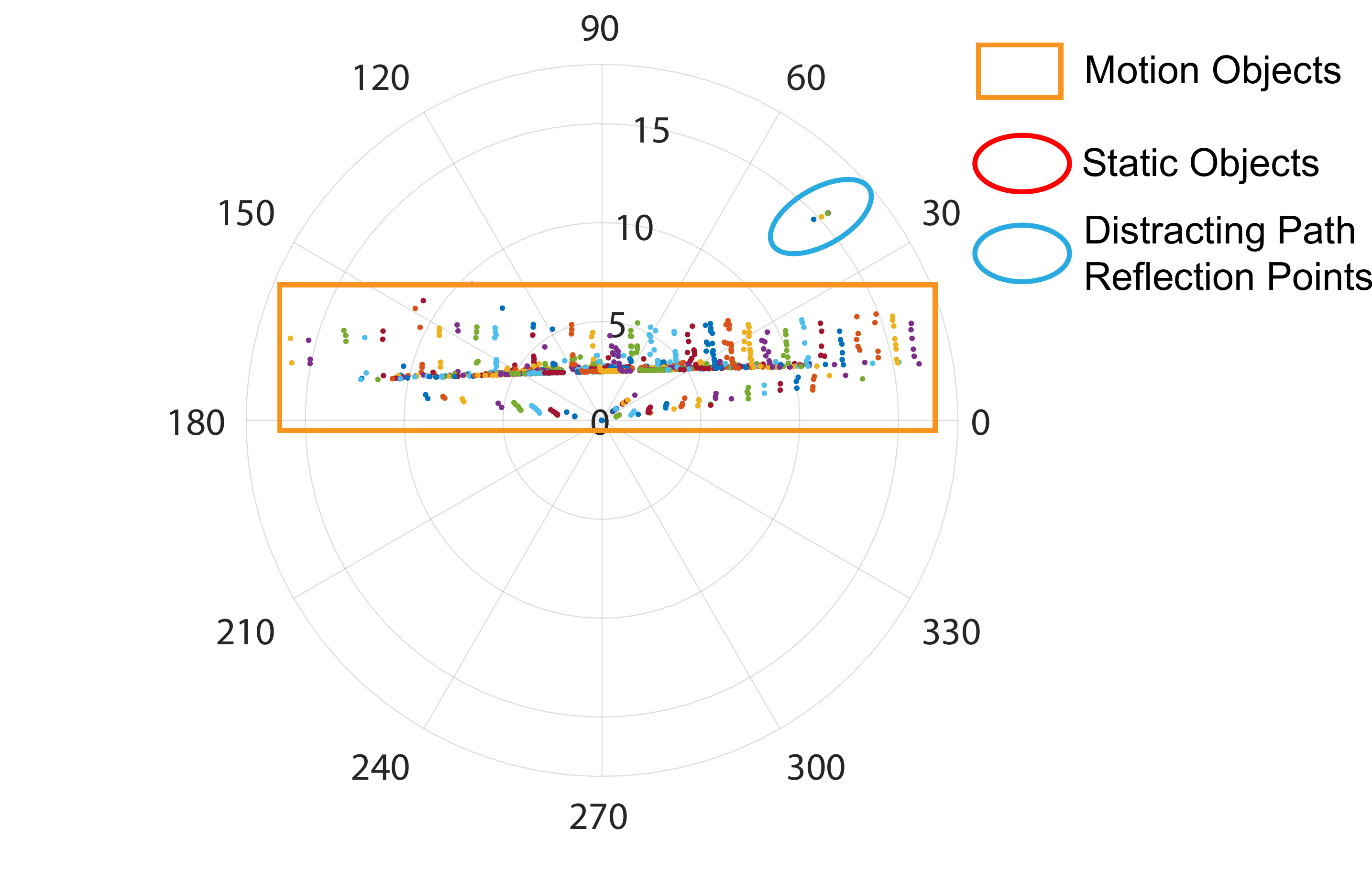}
			\caption{}
			\label{fig:SCR_final}
		\end{subfigure}
		\caption{(a) A moving blockage captured by the RGB camera. (b) The point cloud generated using the raw LiDAR data. It contains the  trace of the moving object, static objects, and distracting path reflection points. (c) The LiDAR  point cloud after  SCR processing. It contains the trace of the moving object and fewer distracting path reflection points.}
		\label{fig:SCR}
	\end{figure*}

	\textbf{Step 3 Distance Quantization:} 
	After angle quantization, the number of important points is $Q$ for each time instance. However, due to the measurement error of the device, the measured distance corresponding to an angle may not be exactly the same from one time instance to the next, so the distance values are also quantized. We choose the total distance quantization levels ($Q_d$) to be 500, the step size is 0.034m which provides sufficient accuracy for our 17m LiDAR range.  
	
	\textbf{Step 4 Dictionary Generation:} 
	For the static cluster dictionary, we choose the samples from $N_d$ time instances ($N_d = 5000$ in our case) which have no moving objects and remove the repeat points.

	\textbf{Step 5 Residual Cluster Removal:} 
	Next, we compare every point in the dataset with every point in the static cluster dictionary. If the point in the dataset is in the static cluster dictionary, this indicates that it corresponds to a static cluster. Then it gets eliminated by assigning 0 to the distance and keeping the angle unchanged. 
	
	\figref{fig:SCR_final} plots the LiDAR point cloud after applying our  static cluster removal algorithm. This figure shows that most of the static clusters are removed and the traces of the moving objects are now clear.

	\section{Experimental Results} \label{sec:exp_results}
	In this section, we describe the development dataset and ML model parameters and then present our experimental results for the LiDAR-aided blockage prediction problem. 
	
	\subsection{Development Dataset and Machine Learning Parameters} \label{subsec:eva_metric}
	
	\textbf{Development Dataset:} We use the seed dataset of scenario 24-27, described in \sref{subsec:scenario}, to construct the development dataset for the LiDAR-aided blockage prediction task following two steps: (i) Constructing the time sequences from the seed dataset. We follow the same footsteps in \cite{mmwave_journal} to extract the time sequences based on the link status labels. We have 1718 sequences, and each sequence has LiDAR data and its corresponding link status. (ii) Generating development dataset for CNN: we use $\mathcal Y_{P} = \{ (\mathcal L_{ob}, b_{T_p})_u \}_{u = 1}^{U}$ to denote the  development dataset. We use \eqref{equ:p1_label} to generate $b_{T_p}$ based on like status, and we apply the sliding window methods \cite{mmwave_journal} to generate $\mathcal L_{ob}$ with time instance length $T_{ob}$. The total number of sequences in this dataset is $U$ = 3436.

	\textbf{CNN Parameter Selection:} 
	The hyper-parameters and parameters of each layer of our CNN model are shown in  \tabref{tbl:CNN_lidar}. After SCR processing, the dimension of LiDAR data changes, resulting in parameter changes in the max pooling layer. We show the parameter format of a CNN layer as input channel - output channel - kernel size - padding, and show the kernel size of the max-pooling layer. All parameters are based on the empirical experiments.
	
	\textbf{Performance Metric:} We adopt the Top-1 accuracy as our main evaluation metric. It is defined as the complement of the prediction error \cite{ImageNet}.

	\begin{table}[tb]
		\centering
		\caption{CNN Model Parameters}
		\begin{tabular}{cccc}
			\cline{1-4}
			\hline\hline
			\multicolumn{2}{c}{\multirow{2}{*}{\textbf{Name}}} & \multicolumn{2}{c}{ \textbf{Value}}  \\ 
			\multicolumn{2}{c}{} &  \textbf{Original}    &  \textbf{SCR}     \\
			\hline
			\multicolumn{2}{c}{Input sequence dimension}       &  16$\times$460$\times$2  &  16$\times$216$\times$2     \\ 
			\multicolumn{2}{c}{Predicted future time steps} &  1-10 &  1-10    \\ 
			\multirow{3}{*}{Stack 1}    & Conv 1   &  2-8-3-1  &  2-8-3-1     \\ 
			& Conv 2   &  8-16-3-1  &  8-16-3-1   \\ 
			&Max pooling 1  &  (2,23)    &  (2,9)      \\  \hline
			\multirow{3}{*}{Stack 2}    & Conv 3   &  16-16-3-1  &  16-16-3-1   \\ 
			& Conv 4   &  16-32-3-1 &  16-32-3-1   \\ 
			& Max pooling 2  &  (2,5)    &  (2,6)     \\ \hline
			\multicolumn{1}{c}{FC}      &            &  (512,2)  &  (512,2)      \\ 
			\multicolumn{1}{c}{Dropout rate}    &    &  0.2  &  0.2   \\ 
			\multicolumn{1}{c}{Epoch}           &    &  1000 &  1000   \\ 
			\multicolumn{1}{c}{Total Parameters} &    &  9306 &  6883   \\

			\hline\hline
		\end{tabular}
		\label{tbl:CNN_lidar}
	\end{table}

	\subsection{Results} \label{subsec:result}

	\begin{figure}[tb]
		\centering
		\includegraphics[width=1\linewidth]{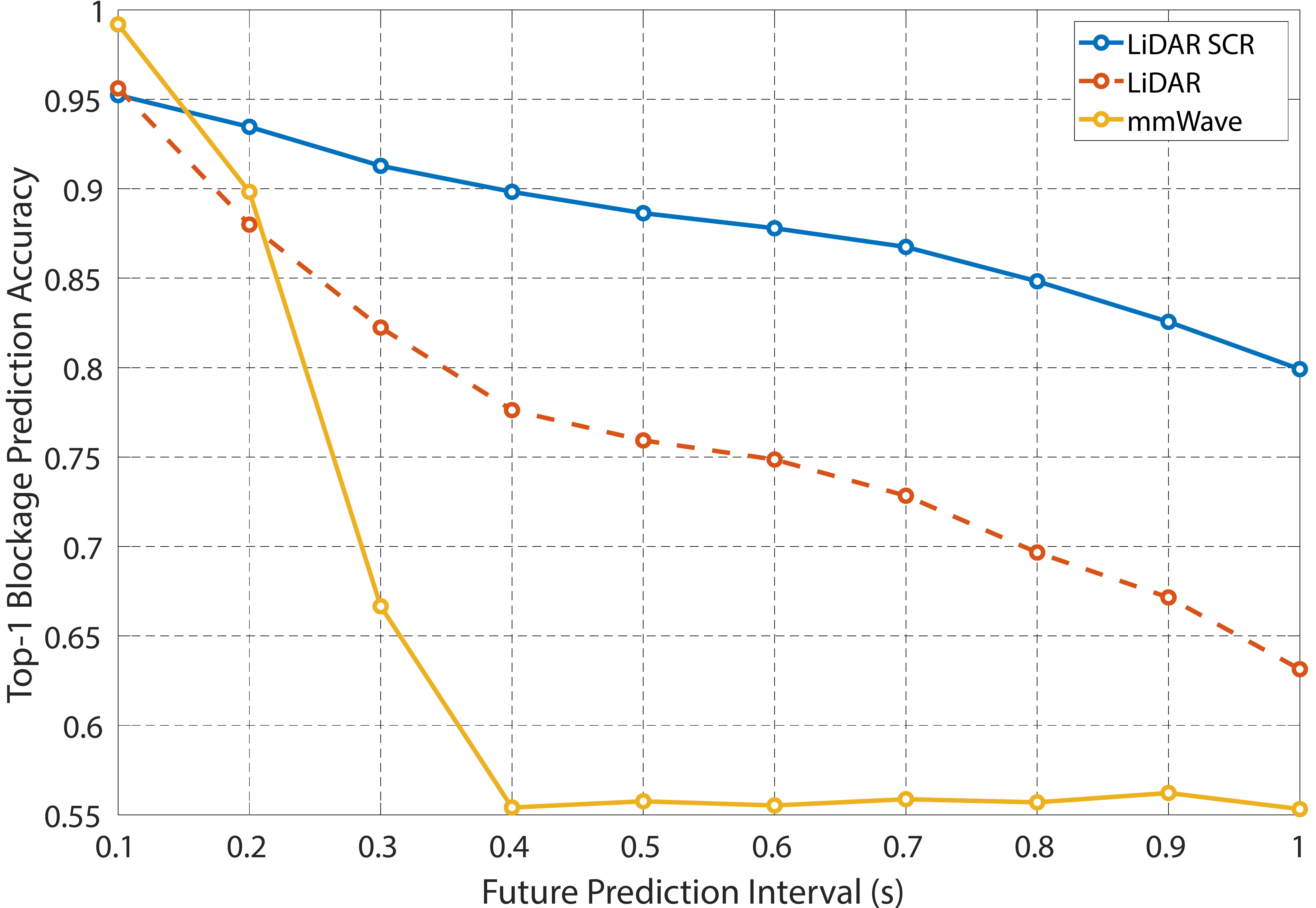}
		\caption{Performance of the proposed blockage occurrence prediction approaches; the LiDAR SCR solution achieves the best performance, espeically for long future prediction intervals.}
		\label{fig:p1_spot0}
	\end{figure}
	
	We evaluate the performance of our blockage prediction model using the LiDAR data and compare it with the mmWave signature based blockage prediction approach in \cite{mmwave_journal}. For the LiDAR data, we use both the original and SCR processed data (using our developed approach in \sref{sec:SCR}). First, we plot the top-1 blockage prediction accuracy versus the future prediction interval (the time before the blockage actually happens) in \figref{fig:p1_spot0}. This figure shows that the LiDAR-aided blockage prediction accuracy greatly improves with the SCR processing. Specifically, \textbf{our model can predict the incoming blockage with more than 80\% accuracy one second before the link is blocked, providing sufficient time for proactive network management.}

	A comparison of the blockage prediction accuracy using our LiDAR data and using the mmWave signatures in \cite{mmwave_journal} is also captured by \figref{fig:SCR}. This figure shows that the approach using wireless signatures outperforms the LiDAR solution when the prediction interval is less than 0.2s. With longer prediction intervals, the wireless signature based solution drops rapidly. This is mainly due to the relatively short (in time) wireless pre-blockage signature compared to the LiDAR pattern. Based on that, the prediction accuracy with the LiDAR data degrades very slowly; even at a prediction interval of 0.6s, the proposed model achieves $\sim 0.75$.

	\begin{figure}[tb]
		\centering
		\includegraphics[width=1.03\linewidth]{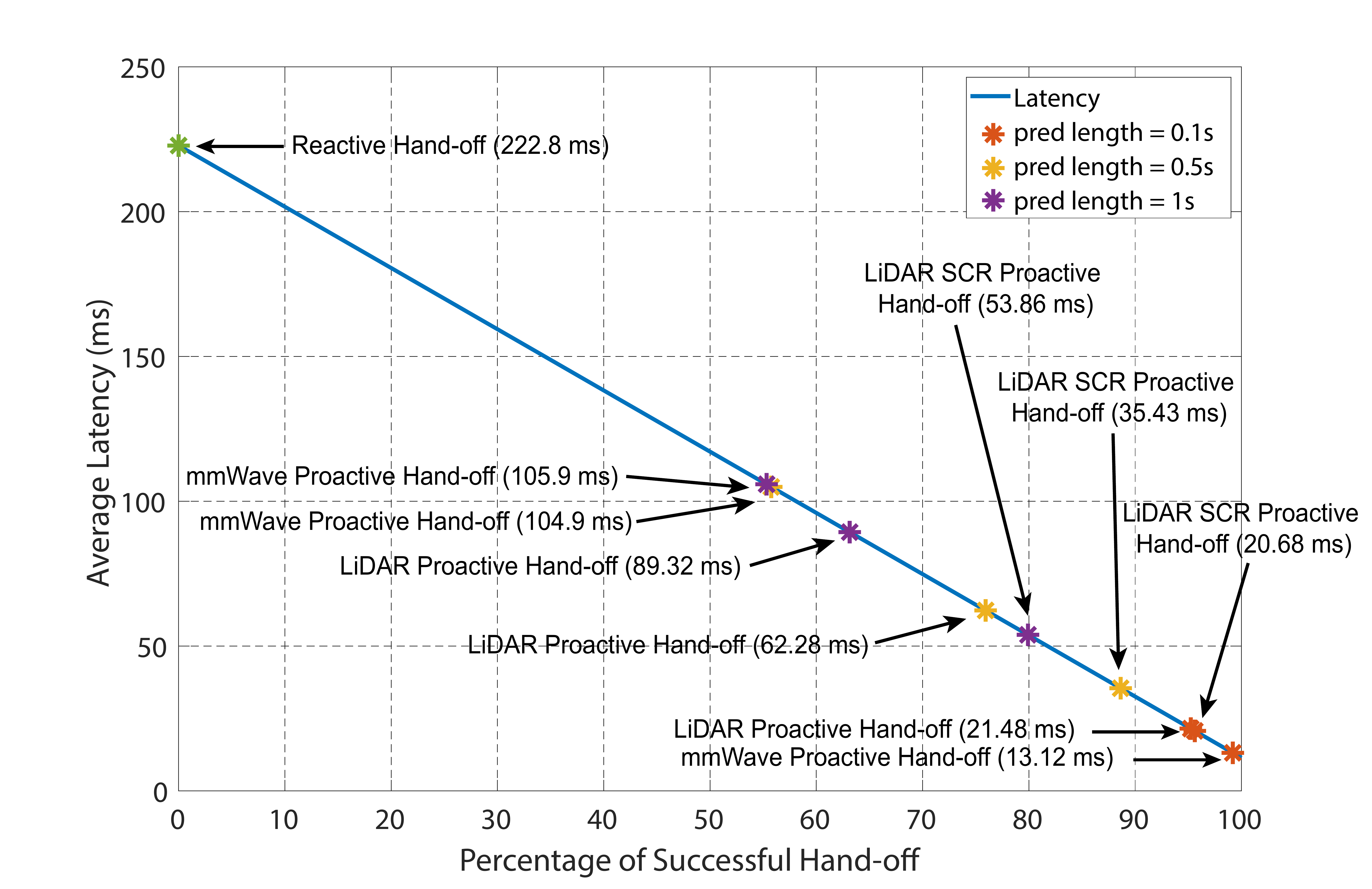}
		\caption{Average latency of the proposed proactive hand-off solutions compared to the conventional reactive approaches.}
		\label{fig:latency}
	\end{figure}
	
	To draw some insights about the potential gains of the proposed methods for the initial access latency in 3GPP 5G NR, we adopt the approach in \cite{charan2021vision} for analyzing the results. According to the 3GPP specifications, a conventional reactive hand-off results in an overall delay of 222.8ms. If the blockage is predicted proactively, a successful proactive hand-off scenario results in 11.4ms latency associated with the contention free random access \cite{charan2021vision}. Based on the prediction accuracy and the latency given by the average latency for the user hand-off, the average latency $\delta$ for the user is given by $\delta = \hat{p} \times 11.4 + (1-\hat{p}) \times 222.8$,  where, $\hat{p}$ is the blockage prediction accuracy at 0.1s, 0.5s and 1s.
	
	\figref{fig:latency} shows the average latency improvement of the proposed methods compared to reactive hand-off. By applying our proposed methods, we can achieve an average latency of 13.12ms using mmWave data, 21.48ms using raw LiDAR data, and 20.68ms using LiDAR SCR data when the prediction length is 0.1s. As the prediction interval increases, the LiDAR SCR based approach consistently maintains low latency compared to the other two solutions. Compared to the reactive hand-off latency, our approaches realize more than 10 times improvement in latency.
	
	\section{Conclusion and Takeaways} \label{sec:con}
	In this paper, we explored the potential of leveraging LiDAR sensory data to proactively predict dynamic blockages in mmWave systems and allowing the network to make proactive management, e.g., hand-off, decisions. We formulated the LiDAR-aided blockage prediction problem and developed an efficient machine learning model for this task based on a CNN architecture. To validate the feasibility of the proposed approach, we constructed a large-scale real-world mmWave-and-LiDAR dataset. Then, we designed a LiDAR data denoising (static cluster removal) algorithm that can enhance the data quality obtained from low-cost LiDAR sensors. Evaluating our developed solutions on this real-world dataset yields the following takeaways: 
	\begin{itemize}
		\item For predicting future moving blockages that are within  a short window (200ms), using the mmWave pre-blockage signature approach in \cite{mmwave_journal} might be sufficient to achieve high accuracy ($>90\%$). 
		\item For predicting moving blockages that are further in the future (up to 1s before they happen), our LiDAR-aided blockage prediction approach achieves more than $80\%$ top-1 accuracy. 
		\item Applying static cluster removal/denoising processing can significantly improve the prediction accuracy, especially with low-cost LiDAR sensors. 
		\item In terms of 3GPP 5G NR latency, the proposed proactive blockage prediction approaches can achieve 10x improvement for the hand-off/beam switching tasks. 
	\end{itemize}
	These results highlight a promising solution for overcoming the blockage challenges in mmWave/THz networks.


\end{document}